\documentclass[secnumarabic,%
aps,%
pra,amsfonts,%
twocolumn,%
showkeys,floatfix%
]{revtex4-1}

\usepackage{bm,amsfonts}
\usepackage{sidecap}
\usepackage{placeins}
\usepackage{multirow}
\usepackage{graphicx}
\newcommand{\gl}[1]{(\ref{#1})}

\newcommand{\lafeaso}{lafeaso1_arxiv}
\newcommand{\enhm}{enhm_arxiv}
\newcommand{\josybacuo}{josybacuo7_mgb2_arxiv}
\newcommand{\josybacuobeide}{josybacuo7_mgb2_ybacuo7_arxiv}
\newcommand{\josn}{josn_arxiv}
\newcommand{\joslacuo}{josla2cuo4_arxiv}
\newcommand{\ybacuo}{ybacuo6_arxiv}
\newcommand{\josm}{josm_arxiv}

\renewcommand{\thetable}{\arabic{table}}

\begin{document}
\title{The reason why doping causes superconductivity in LaFeAsO}  
\author{Ekkehard Kr\"uger}
\author{Horst P. Strunk}
\affiliation{Institut f\"ur Materialwissenschaft, Materialphysik,
  Universit\"at Stuttgart, D-70569 Stuttgart, Germany}
%
\date{\today}
\begin{abstract}
  The experimental observation of superconductivity in LaFeAsO appearing on
  doping is analyzed with the group-theoretical approach that evidently led in
  a foregoing paper (J. Supercond 24:2103, 2011) to an understanding of the
  cause of both the antiferromagnetic state and the accompanying structural
  distortion in this material.  Doping, like the structural distortions, means
  also a reduction of the symmetry of the pure perfect crystal.  In the
  present paper we show that this reduction modifies the correlated motion of
  the electrons in a special narrow half-filled band of LaFeAsO in such a way
  that these electrons produce a stable superconducting state.
\end{abstract}
\maketitle

\section{Introduction}
Undoped LaFeAsO does not become superconducting at any temperature but rather
undergoes a structural distortion from tetragonal to orthorhombic symmetry at
$\sim\! 155K$ as well as an antiferromagnetic spin ordering transition at
$\sim\!  137K$ \cite{clarina, nomura,kitao,nakai}. As oxygen is gradually
replaced with fluorine, however, the antiferromagnetic order in the doped
material LaFeAsO$_{1-x}$F$_x$ disappears and a superconducting state occurs at
the relatively small doping level $x \approx 0.05$. When $x$ continues to
increase the orthorhombic structural distortion coexists with
superconductivity until the superconducting phase becomes tetragonal at doping
levels $x \gtrsim 0.08$~\cite{huang,margadonna}.  On the other hand, there is
no experimental evidence of the antiferromagnetic order coexisting with the
superconducting state.

The evidently successful foundation of the existence of both the
antiferromagnetic state and the accompanying structural distortion in LaFeAsO
given in a foregoing paper~\cite{\lafeaso} has led us to consider now the
superconducting state and the orthorhombic non-magnetic distortion of LaFeAsO
using the same approach. Though this approach is unusual in the
superconductivity field, it corroborates the generally accepted idea that {\em
  correlated} electrons are responsible for the superconducting state. It is
based on a group theoretical treatment of correlated atomic-like electrons
(Sec.~\ref{sec:nhm}) in narrow, half-filled ``superconducting bands'' of
special symmetry as they are considered within a nonadiabatic extension of the
Heisenberg model, the nonadiabatic Heisenberg model \cite{\enhm} (NHM). We
will put also emphasis on making clear the underlying physical ideas of this
model.

In Sec.~\ref{sec:superconductivity} we sketch the central role of
superconducting bands (as they shall be defined in the
Appendix~\ref{sec:definition}) in Cooper pair formation within the NHM. Then,
in Sec.~\ref{sec:nonexistence}, we show that tetragonal undoped LaFeAsO (with
the space group
$P4/nmm$~\cite{clarina,kamihara,prl_chen,nature_chen,wen,dong}) does not
possess a superconducting band. However, a superconducting band does appear in
the band structure of LaFeAsO when the symmetry of the material is reduced. As
is shown, this reduction can be realized by the doping
(Sec.~\ref{sec:dopedmat}) or by an orthorhombic distortion of the crystal
(Sec.~\ref{sec:orthorhombic}).

\section{Superconducting Bands and Superconductivity}
\label{sec:superconductivity}
In this section we outline the mechanism of Cooper pair formation within the
NHM, for more detailed summaries see Ref.~\cite{\josybacuo} or~\cite{\josn}.

\subsection{Nonadiabatic Heisenberg model}
\label{sec:nhm}
The NHM was defined in Ref.~\cite{\enhm}. It is bases on three postulates
concerning the {\em atomic-like} motion of the electrons in narrow,
half-filled energy bands as it was first considered by Mott \cite{mott} and
Hubbard\cite{hubbard}: the electrons occupy localized states as long as
possible and perform their band motion by hopping from one atom to
another. The three postulates of the NHM are physically evident, emphasize the
{\em correlated} nature of any atomic-like motion, and require the
introduction of {\em nonadiabatic} localized states. In this way the NHM
refines the original treatment of atomic-like electrons as it was given by
Mott and Hubbard. Fortunately, only the symmetry of the localized functions
representing the nonadiabatic localized states need to be known. Within the
NHM they have the same symmetry and spin dependence as the best localized {\em
  exact} Wannier functions of the considered partly filled energy band. In
this context, ``exact'' Wannier functions form a {\em complete} basis of the
Bloch functions that {\em really exist} in the band.

An important feature of the NHM is the existence of the nonadiabatic
condensation energy $\Delta E$ as defined in Eq.~(2.20) of Ref.~\cite{\enhm}:
the electron system gains $\Delta E$ at the transition from a purely bandlike
state to the correlated (nonadiabatic) atomic-like state. This energy gain is
caused by the Coulomb {\em correlation} energy and, thus, the NHM yields an
important and novel contribution to the theory of correlated electrons.

\subsection{New mechanism of Cooper pair formation}

In case the considered energy band is a superconducting band, the related
Wannier functions are {\em spin-dependent}, see the detailed definition given
in the Appendix~\ref{sec:definition}. The symmetry of these Wannier functions
ensures that the nonadiabatic Hamiltonian $H^n$ commutes correctly with the
symmetry operations of the space group. Their spin dependence, on the other
hand, has the far reaching consequence that only in special atomic-like states
the electrons satisfy the conservation of spin angular-momentum. For this
reason the Coulomb interaction in a narrow, roughly half-filled
superconducting band produces an interaction between the electron spins and
``crystal-spin-1 bosons''. These crystal-spin-1 bosons are the {\em
  energetically lowest} boson excitations of the crystal that possess the
crystal-spin angular momentum $1\cdot\hbar$ and are sufficiently stable to
transport it through the crystal. 

 \begin{figure*}[t]
  \includegraphics[width=.85\textwidth,angle=0]{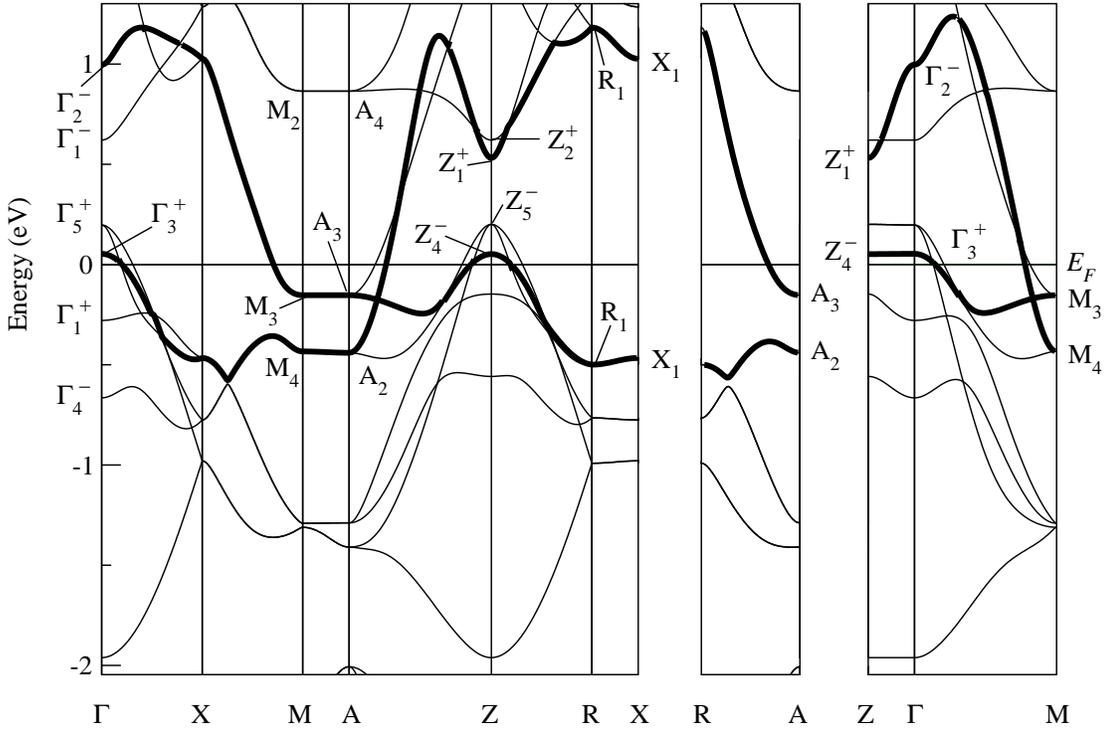}
  \caption{ Band structure of tetragonal LaFeAsO as calculated by the FHI-aims
    program \protect\cite{blum1,blum2}, with symmetry labels determined by the
    authors. The symmetry labels can be identified from Table A.1 of
    Ref~\protect\cite{\lafeaso}. The bold line shows the active band that
    becomes a superconducting band (as defined in the
    Appendix~\ref{sec:definition}) in the doped material. 
  \label{fig:bandstr}
}
 \end{figure*}


Below a certain transition temperature $T_c$, this spin-boson interaction {\em
  forces} the electrons in a novel way to form Cooper pairs. On the one hand,
the crystal-spin-1 bosons mediate the formation of Cooper pairs as it is
already described in the classical~\cite{bcs} theory of superconductivity. In
addition, however, the crystal-spin-1 bosons in a narrow superconducting band
produce {\em constraining forces} as they are familiar from classical
mechanics: below $T_c$, they reduce the degrees of freedom of the electron
system by forcing the electrons to form pairs that are invariant under time
inversion, i.e., by forcing the electrons to form Cooper pairs. A comparison
of the quantum system with a classical system suggests that these constraining
forces operating in a narrow, roughly half-filled superconducting band are
{\em required} for the Hamiltonian of the system to possess {\em eigenstates}
in which the electrons form Cooper pairs. Thus, the fundamental contribution
of the NHM to the theory of superconductivity is the assertion that only
crystal-spin-1 bosons in a narrow superconducting band are able to produce
Cooper pairs.  The NHM predicts that {\em a material can become
  superconducting only if this material has a narrow, roughly half-filled
  superconducting band in its band structure.} This fundamental statement of
the NHM is already corroborated by numerous superconductors and
non-superconductors, see Sec.~\ref{sec:conclusions}.

Constraining forces do not alter the energy of the electron system but only
lower the degrees of freedom. Thus, the vast majority of the statements and
calculations of the traditional theory of superconductivity {\em should stay
  valid} in a superconducting band.

\section{Superconductivity forbidden in pure perfect 
  L\lowercase{a}F\lowercase{e}A\lowercase{s}O}
\label{sec:nonexistence}
Consider the energy band denoted in Fig.~\ref{fig:bandstr} by the bold lines
which in the following shall be called the ``active band'' of LaFeAsO.  In
this section we show that neither this active band nor any other roughly
half-filled energy band of perfect LaFeAsO is a superconducting band.

\begin{SCfigure*}[1][!]
\begin{minipage}[t]{.6\textwidth}
\centering
 \includegraphics[width=.46\textwidth,angle=0]{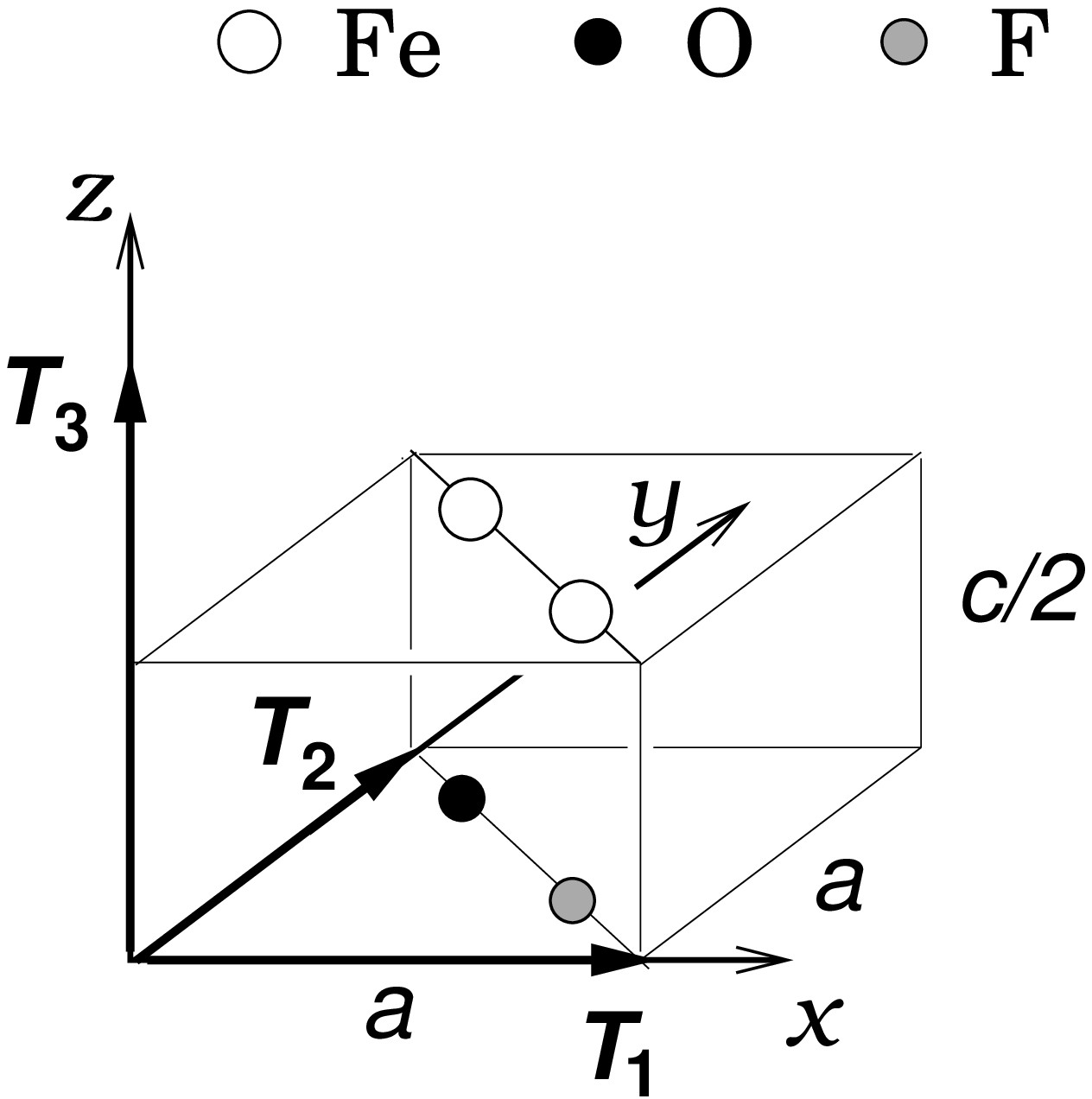}%
\begin{center}
(a)
\vspace{1cm}
\end{center}
\includegraphics[width=.54\textwidth,angle=0]{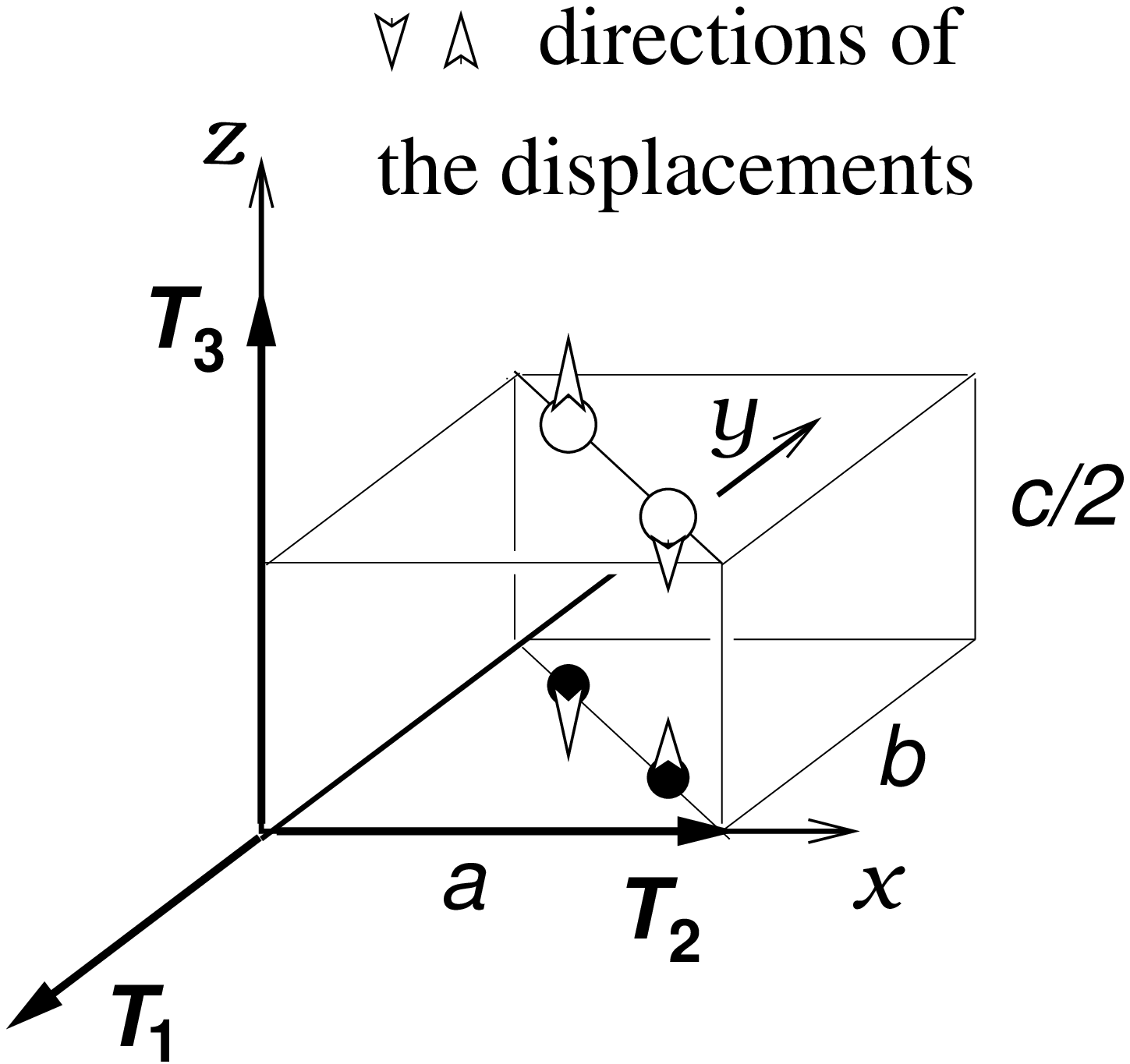}%
\begin{center}
(b)
\vspace{1cm}
\end{center}
\end{minipage}
\columnsep = -20cm
\caption{
Coordinate systems and unit cells of tetragonal (a) and orthorhombically distorted (b)
non-magnetic LaFeAsO. For reasons of clarity, only the iron and oxygen atoms are
shown, where
one of the two oxygen atoms in (a) is replaced by fluorine.  $a$, $b$, and $c$ 
denote the lengths of the sides in the unit cells. The coordinate systems
define the symmetry operations 
$\{R|pqr\}$ as used in this paper. They are written in the Seitz notation
as detailed in the textbook
  of Bradley and Cracknell \protect\cite{bc}: $R$ stands for a point group operation
  and $pqr$   
  denotes the subsequent translation. The point group operation $R$ is related 
  to the $x, y, z$ coordinate system and $pqr$ stands for the translation 
  $\bm t = p\bm T_1 + q\bm T_2 + r\bm T_3$. As defined in Table 3.1 of
  Ref.~\protect\cite{bc}, the basic translations $\bm T_i$ have different
  directions in the 
  tetragonal and orthorhombic structure.  
  The depicted displacements of the Fe and O atoms in (b) realize the
  orthorhombic space
  group $Pmm2$ (25) which allows the onset of superconductivity, see
Sec.~\protect\ref{sec:dopedmat}. The Fe and O atoms are displaced in $\pm z$
direction where the amount of the
displacements of the Fe atoms or of the O atoms in the unit cell is
different in order to distinguish $Pmm2$ from the space group $Pmmn$ (59). 
\label{fig:structures}
}
\end{SCfigure*}


First, consider spin-dependent Wannier functions that are situated on the Fe
sites. Table~\ref{tab:wf129} (b) lists the double-valued representations of
all the superconducting bands in tetragonal undoped LaFeAsO with the
spin-dependent Wannier functions situated on the Fe sites. Each of these two
superconducting bands consists of two branches because there are two Fe atoms
in the unit cell. Table~\ref{tab:wf129} (b) shows that the two branches of
either superconducting band are degenerate at the points $M$ and $A$ because,
in each of these points, the Bloch functions belong only to one
(double-valued) representation, namely to the {\em four-}dimensional
representation $M_5$ and $A_5$, respectively

The two branches of the active band of LaFeAsO, on the other hand, are split
at points $M$ and $A$. Thus, it is not identical with one of the
superconducting bands listed in Table~\ref{tab:wf129} (b). Even if we allow
the related spin-dependent Wannier functions to be centered at the La, As, or
O atoms, the active band does not become a superconducting band because the
two branches of any superconducting band in tetragonal undoped LaFeAsO are
degenerate at points $M$ and $A$. This important feature follows from the mere
fact that both points $M$ and $A$ possess in the space group $P4/nmm$ only
{\em one} doubled-valued representation, namely the four-dimensional
representation $M_5$ and $A_5$, respectively, see Table 6.13 of
Ref.~\cite{bc}.

Consequently, the active band is not a superconducting band. We can neither
combine the bands near the Fermi level in the band structure of LaFeAsO in any
other way to form a continuous superconducting band. Any roughly half-filled
energy band in tetragonal undoped LaFeAsO (consisting of two branches) which
is compatible with either superconducting band in Table~\ref{tab:wf129} (b)
would have a considerable gap between $M$ and $X$ and between $A$ and $R$,
which extends over the whole plane $M X R A$. Such a strongly discontinuous
energy band, however, does not yield best localized spin-dependent Wannier
functions and, hence, is unsuitable to represent atomic-like electrons.

To sum up, in the band structure of perfect LaFeAsO a superconducting band
does not exist, and thus superconductivity cannot develop.

\section{Superconductivity in weakly doped
  L\lowercase{a}F\lowercase{e}A\lowercase{s}O}
\label{sec:dopedmat}
\subsection{The space group $P\overline{4}m2$ of weakly doped LaFeAsO}
\label{sec:spacegroup}
The doped material LaFeAsO$_{1-x}$F$_x$ (with $x > 0$) no longer possesses any
space group symmetry because the doping destructs the translation symmetry of
the crystal. Thus, the electrons no longer move in a potential invariant under
the tetragonal space group $P4/nmm$. However, even in the doped material the
array of the Fe atoms is invariant under the translations of
$P4/nmm$. Therefore we describe approximately for small doping levels $x$ the
atomic-like state related to the Fe atoms by an idealized atomic-like state in
a potential which still is invariant under a tetragonal space group $S$. The
best choice of this space group $S$ is that subgroup of $P4/nmm$ which is {\em
  best adapted} to the local symmetry of the oxygen-fluorine layers.  Thus, we
get the optimal choice of the group $S$ by removing from $P4/nmm$ all the
(proper and improper) rotations and reflections not leaving invariant the
positions of the fluorine atoms. In this way we obtain the group $S =
P\overline{4}m2 = \Gamma_qD^5_{2d}$ (115) still possessing the
tetragonal-primitive Bravais lattice $\Gamma_q$.

Hence, we use the group $P\overline{4}m2$ as the space group of the
atomic-like electrons centered on the Fe sites in weakly doped LaFeAsO. This
group can be defined by the two generating elements
\begin{equation}
  \label{eq:9}
\{S^+_{4z}|0\textstyle\frac{1}{2}0\}\ \text{and}\ 
\{C_{2a}|\textstyle\overline{\frac{1}{2}}\textstyle\frac{1}{2}0\},
\end{equation}
cf. Table 3.7 of Ref.~\cite{bc} and Table~\ref{tab:gruppen}. The symmetry
operations~\gl{eq:9} are given in the coordinate system defined in
Fig.~\ref{fig:structures} (a). The generating elements define a group because
every element of the group is expressible as a product of powers of the
generating elements. Hence, a given structure is invariant under all the
symmetry operations of a space group if it is invariant under the generating
elements.

\begin{table*}[!]
\caption{Space groups of LaFeAsO as determined in this paper and in
  Ref.~\protect\cite{\lafeaso}.  
\label{tab:gruppen}}
\begin{tabular}[t]{llllcclc}
\\
\multicolumn{1}{l}{Inter-}&
\multicolumn{1}{l}{Sch\"onflies}&
\multicolumn{1}{l}{Inter-}&
\multicolumn{1}{l}{Coordinate}&
\multicolumn{1}{c}{$\bm t_{bc}$}&
\multicolumn{1}{c}{$\bm t_{int}$}&
\multicolumn{1}{c}{Descripion}&
\multicolumn{1}{c}{Doping}\\
\multicolumn{1}{l}{nat.}&
\multicolumn{1}{l}{symbol}&
\multicolumn{1}{l}{nat.}&
\multicolumn{1}{l}{system and}\\
\multicolumn{1}{l}{symbol}&
\multicolumn{1}{l}{}&
\multicolumn{1}{l}{num-}&
\multicolumn{1}{l}{displace-}&&&
\multicolumn{1}{c}{(cf. Fig.~\protect\ref{fig:gruppen})}\\
\multicolumn{1}{l}{}&
\multicolumn{1}{l}{}&
\multicolumn{1}{l}{ber}&
\multicolumn{1}{l}{ments of}\\
\multicolumn{1}{l}{}&
\multicolumn{1}{l}{}&
\multicolumn{1}{l}{}&
\multicolumn{1}{l}{Fe and O}\\
\hline
$Imma$ &  $\Gamma_o^vD^{28}_{2h}$ & 74 & Fig.~3 (b) &
$\frac{1}{2}\bm T_3$ &
$\frac{3}{4}\bm T_1 + \frac{1}{4}\bm T_2 + \frac{1}{2}\bm T_3$
&
Hypothetical antiferromagnetic &  undoped\\
&&&of Ref.~\protect\cite{\lafeaso} &&&
structure in undistorted LaFeAsO  &\\
$Pnn2$ & $\Gamma_oC^{10}_{2v}$ &  34 & Fig.~3 (c) &
$\frac{1}{4}\bm T_2$ & $-\frac{1}{4}\bm T_1 + \frac{1}{2}\bm T_2$ &
Antiferromagnetic structure in 
 &  undoped\\
&&&of Ref.~\protect\cite{\lafeaso} &&&
distorted LaFeAsO&\\
$P\overline{4}m2$ & $\Gamma_qD^5_{2d}$ & 115 & Fig.~\ref{fig:structures} (a) &
$\frac{1}{4}\bm T_1 - \frac{1}{4}\bm T_2$ & $\frac{1}{4}\bm T_1 -
\frac{1}{4}\bm T_2$
&
Tetragonal LaFeAsO including the & weakly doped\\ 
&&&&&&
tetragonal superconducting phase &\\ 
$Pmm2$ & $\Gamma_oC^1_{2v}$ & 25 & Fig.~\ref{fig:structures} (b) &
$\frac{1}{4}\bm T_1 + \frac{1}{4}\bm T_2$ &  
$\frac{1}{4}\bm T_1 + \frac{1}{4}\bm T_2$&  
Non-magnetic orthorhombically & undoped or\\
&&&&&&  
distorted LaFeAsO including the & weakly doped\\
&&&&&&  
orthorhombic superconducting phase &\\
\hline\\
\end{tabular}\hspace{1cm}
\ \\
\begin{flushleft}
Notes to Table~\ref{tab:gruppen}
\end{flushleft}
\begin{enumerate}
\item $Imma$ is the space group of the experimentally observed
  antiferromagnetic structure in LaFeAsO as given in Fig.~4 of
  Ref.~\protect\cite{clarina}. We call it ``hypothetical'' because it is
  unstable in undistorted LaFeAsO.
\item In any case, the orthorhombic distortion of LaFeAsO is realized by the
  indicated displacements of the Fe and O atoms {\em alone}.
\item The origin of the coordinate system used in Ref.~\protect\cite{bc} and
  in the {\em International Tables for Crystallography} is translated into the
  origin used in this paper by the indicated vectors $\bm t_{bc}$ and $\bm
  t_{int} = \bm t_{bc} + \bm t_0$, respectively, where $\bm t_0$ is given in
  Table 3.7 of Ref.~\protect\cite{bc}. Hence, the space group elements $P_p$
  used in this paper may be determined by the equations $P_{p} = \{E|\bm
  t_{bc}\}P_{bc}\{E|-\bm t_{bc}\}$ and $P_{p} = \{E|\bm
  t_{int}\}P_{int}\{E|-\bm t_{int}\}$ from the space group elements $P_{bc}$
  and $P_{int}$ of Ref.~\protect\cite{bc} and of the {\em International
    Tables}, respectively, where $E$ denotes the indentity operation.
\end{enumerate}
\end{table*}



\subsection{Superconducting band in the space group $P\overline{4}m2$}
\label{sec:superconductingband}
In this section we show that the active band of LaFeAsO becomes a
superconducting band in the space group $P\overline{4}m2$ of doped
LaFeAsO. For this purpose we unitarily transform the Bloch functions of the
active band into Bloch functions adapted to the symmetry of the space group
$P\overline{4}m2$, while the run of the active band in the band structure of
LaFeAsO is assumed to be not considerably affected by the doping.

Table~\ref{tab:wf115} lists all the energy bands in LaFeAsO whose Bloch
functions can be unitarily transformed into best localized Wannier functions
[Table~\ref{tab:wf115} (a)] or spin-dependent Wannier functions
[Table~\ref{tab:wf115} (b)] which are centered at the Fe atoms and
are symmetry-adapted to the space group $P\overline{4}m2$.

Comparing the bands listed in Table~\ref{tab:wf129} with the bands given in
Table~\ref{tab:wf115}, we see the important difference between these bands:
while the two branches of all the bands in Table~\ref{tab:wf129} are
degenerate at points $M$ and $A$, the two branches of the bands given in
Table~\ref{tab:wf115} are not degenerated at any point of symmetry in the
Brillouin zone since at each point of symmetry $\Gamma, M, Z, A, R, \text{ and
}X$ the Bloch states belong to a sum of two representations. Thus, in the
space group $P\overline{4}m2$ the active band of LaFeAsO may be identical with
one of the bands listed in Table~\ref{tab:wf115} since the active band as well
as the bands listed in Table~\ref{tab:wf115} are nowhere degenerate. We now
show that indeed the active band is almost identical with band 4 in
Table~\ref{tab:wf115} (a) and fully identical with band 2 in
Table~\ref{tab:wf115} (b).

Within the space group $P4/nmm$, the Bloch functions of the active band are
labeled by the representations
\begin{equation}
\label{eq:1}
\begin{array}{llllll}
\Gamma^-_2,& \Gamma^+_3; \ &
M_3,& M_4; \ &
A_3,& A_2;\\
Z^+_1,& Z^-_4; \ &
R_1,& R_1; \ &
X_1,& X_1.\\
\end{array}
\end{equation}
According to Table~\ref{tab:faltena}, we may unitarily transform these
functions into Bloch functions labeled by the representations of the space
group $P\overline{4}m2$ (possessing the same Brillouin zone $\Gamma_q$),
\begin{equation}
  \label{eq:2}
\begin{array}{lclp{.3cm}lcl}
\Gamma^-_2 & \rightarrow & \underline{\Gamma_4}, && 
\Gamma^+_3 & \rightarrow & \underline{\Gamma_4};\\[.2cm] 
M_3 & \rightarrow & M_5,&&  
M_4 & \rightarrow & \underline{M_2} + M_4;\\[.2cm]
A_3 & \rightarrow & A_5,&&  
A_2 & \rightarrow & \underline{A_1} + A_3;\\[.2cm]
Z^+_1 & \rightarrow & \underline{Z_1},&&  
Z^-_4 & \rightarrow & \underline{Z_1};\\[.2cm]  
R_1 & \rightarrow & \underline{R_1} + R_4,&&  
R_1 & \rightarrow & R_1 + \underline{R_4};\\[.2cm]  
X_1 & \rightarrow & \underline{X_1} + X_4,&&  
X_1 & \rightarrow & X_1 + \underline{X_4}.\\[.2cm]  
\end{array}
\end{equation}
The underlined representations belong to a band listed in
Table~\ref{tab:wf115}, namely to band 4 in Table~\ref{tab:wf115} (a). The two
representations $M_5$ and $A_5$, however, are {\em not} part of band 4. Thus,
we {\em cannot} unitarily transform the Bloch functions of the active band
into usual (i.e., spin-{\em in}dependent) Wannier functions being best
localized, centered at the Fe atoms, and symmetry-adapted to the group
$P\overline{4}m2$. [The single-valued representations $M_5$ and $A_5$ of
$P\overline{4}m2$ listed in~\gl{eq:2} should not be confused with the double-valued
representations $M_5$ and $A_5$ of $P4/nmm$ considered in
Sec.~\ref{sec:nonexistence}].

The situation is changed when we allow the Wannier functions to be
spin-dependent. In this case we may replace the single-valued representations
$R_i$ by the corresponding {\em double-valued} representations $R_i\times
D_{1/2}$, where $D_{1/2}$ denotes the two-dimensional double-valued
representation of the three-dimensional rotation group $O(3)$ given, e.g., in
Table 6.1 of Ref.~\cite{bc}. Table~\ref{tab:faltenb} lists all the
representations $R_i\times D_{1/2}$ in the space group $P\overline{4}m2$. In
particular, it shows that the two unsuitable representations $M_5$ and $A_5$
in \gl{eq:2} split at the transition from the single-valued to the
double-valued representations,
\begin{equation}
  \label{eq:3}
\begin{array}{lcl}
M_5\times D_{1/2}& = &M_6 + M_7,\\  
A_5\times D_{1/2}& = &A_6 + A_7.
\end{array}
\end{equation}
For that reason, the active band becomes identical with band 2 in
Table~\ref{tab:wf115} (b). Consequently, the Bloch functions of the active
band can be unitarily transformed into {\em spin-dependent} Wannier functions
being best localized, centered at the Fe atoms, and symmetry-adapted to the
group $P\overline{4}m2$. These Wannier functions are only {\em weakly}\
spin-dependent because it is only the two representations $M_5$ and $A_5$ in
\gl{eq:2} which are not part of band 4 in Table~\ref{tab:wf115} (a). This
observation together with former observations on
La$_2$CuO$_4$~\cite{\joslacuo}, YBa$_2$Cu$_3$O$_7$, and
MgB$_2$~\cite{\josybacuobeide} suggests that this weak spin dependence is an
additional condition for stable high-$T_c$ superconductivity. 

To sum up, within the space group $P\overline{4}m2$ the active band of LaFeAsO
is a superconducting band related to the Fe atoms. Consequently, in the doped
material the electron system may gain the nonadiabatic condensation energy
$\Delta E$ by producing the special spin-boson interaction acting in a
superconducting band. Due to this interaction, the electrons are forced to
form Cooper pairs below a certain transition temperature. The resulting
superconducting state is experimentally well-established~\cite{kamihara}.

\section{Superconductivity in distorted
  L\lowercase{a}F\lowercase{e}A\lowercase{s}O}
\label{sec:orthorhombic}
The doping reduces the symmetry of pure LaFeAsO in such a way that the active
band of LaFeAsO becomes a superconducting band (preceding
Sec.~\ref{sec:dopedmat}). However, the same effect may have a reduction of the
symmetry produced by a spatial distortion of undoped LaFeAsO. There are
several subgroups of the space group $P4/nmm$ of tetragonal undoped LaFeAsO
allowing a stable superconducting state. In view of the experimental results
\cite{clarina} on LaFeAsO, we restrict ourselves to the distortion of the
lattice produced by the (small) displacements of the Fe and O atoms depicted
in Fig.~\ref{fig:structures} (b). Depending on whether or not the unit cell of
LaFeAsO is doubled in $z$ direction, these displacements realize the
orthorhombic space group $Pmm2$ (25) or the tetragonal space group $P4_2mc$
(105).

\subsection{The orthorhombic space group $Pmm2$ (25) with single unit cell} 

Removing from the space group $P4/nmm$ all the symmetry operations not leaving
invariant the displaced Fe and O atoms as depicted in
Fig.~\ref{fig:structures} (b), we obtain the orthorhombic space group $Pmm2$
defined by the two generating elements
\begin{equation}
    \label{eq:4}
\{\sigma_x|0\textstyle\frac{1}{2}0\}\ \text{and}\ 
\{\sigma_y|\textstyle\frac{1}{2}00\},
\end{equation}
as given in Table 3.7 of Ref.~\cite{bc}, which in this paper are transformed
into the coordinate system defined by Fig.~\ref{fig:structures} (b), see
Table~\ref{tab:gruppen}.

A band identified as superconducting band in the space group $P\overline{4}m2$
is a superconducting band in any subgroup of $P\overline{4}m2$, too. To show
that within our coordinate systems $Pmm2$ is a subgroup of $P\overline{4}m2$,
we write the translations of the generating elements in Eq.~\gl{eq:4} in terms
of the basic translations of the Bravais lattice $\Gamma_q$ of
$P\overline{4}m2$ as depicted in Fig.~\ref{fig:structures} (a). So we get the
symmetry operations $\{\sigma_x|\textstyle\frac{1}{2}00\}$ and
$\{\sigma_y|0\textstyle\overline{\frac{1}{2}}0\}$, which in fact are elements
of $P\overline{4}m2$, see the symmetry operations belonging to point $\Gamma$
in Table~\ref{tab:rep115}. Hence, the active band of LaFeAsO is a
superconducting band in the orthorhombically distorted material with the space
group $Pmm2$. Consequently, the displacements of the Fe and O atoms depicted
in Fig.~\ref{fig:structures} (b) enable the electron system to gain the
nonadiabatic condensation energy $\Delta E$ by occupying an atomic-like state
producing the interaction of the electron spins with crystal-spin-1 bosons
leading to superconductivity.

The fact that $Pmm2$ is a subgroup of $P\overline{4}m2$ has another important
consequence: the orthorhombic structural distortion with the space group
$Pmm2$ is not destroyed by the doping. That means that a superconducting state
enabled by the orthorhombic distortion can coexist with the doping, as it is
experimentally observed~\cite{huang}.

\subsection{The tetragonal space group $P4_2mc$ (105) with doubled unit cell} 

The displacements of the Fe and O atoms as depicted in
Fig.~\ref{fig:structures} (b) realize a further subgroup of $P4/nmm$ if the
unit cell of LaFeAsO is doubled in $z$ direction. In this case, the
displacements no longer are periodic with $\bm T_3$ as given in
Fig.~\ref{fig:structures} (b) but a translation by $\bm T_3$ effects an
inversion of the displacements as depicted in Fig.~3 (c) of
Ref.~\cite{\lafeaso}.

These displacements of the Fe and O atoms with doubled unit cell realize the
space group $P4_2mc$ which would allow a stable superconducting state,
too. However, $P4_2mc$ is tetragonal which contradicts the experimental
observation that the non-magnetic distortion observed in
LaFeAsO~\cite{clarina, nomura,kitao,nakai} is orthorhombic. In addition,
$P4_2mc$ is not a subgroup of the space group $P\overline{4}m2$ of the weakly
doped material. Consequently, if $P4_2mc$ would be realized in LaFeAsO, we
could hardly understand the experimental observation~\cite{huang} that the
non-magnetic distortion can survive in the doped system. So we propose that
the non-magnetic distortion of LaFeAsO possesses the orthorhombic space group
$Pmm2$ rather than the tetragonal space group $P4_2mc$.

\section{Concept of correlated electrons in pure and doped
  L\lowercase{a}F\lowercase{e}A\lowercase{s}O}
\label{sec:concept}
\begin{figure}[t]
 \includegraphics[width=.45\textwidth,angle=0]{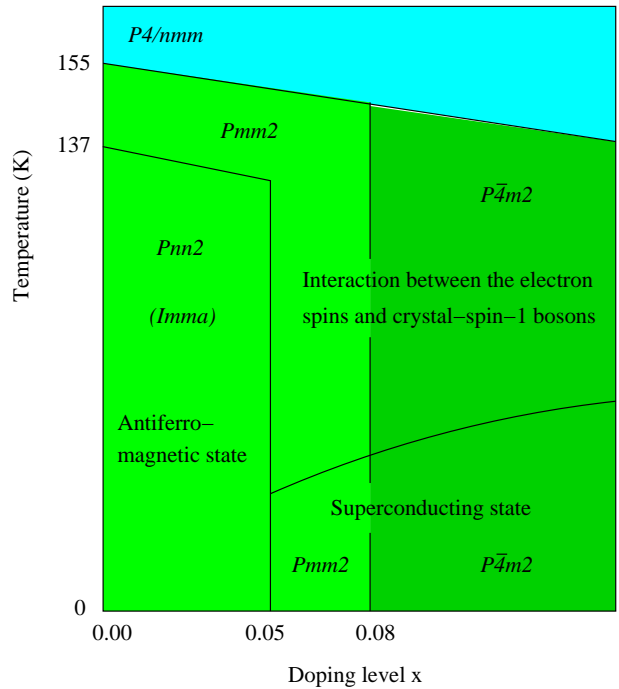}
 \caption{
   Schematic structural phase diagram of the space groups listed in
   Table~\ref{tab:gruppen} as  
   suggested by our results and experimental observations~\protect\cite{clarina,
     nomura,kitao,nakai,huang,margadonna}. In the blue region the electrons
   perform a purely bandlike motion and in the green region the electrons of the
   active band occupy an atomic-like state as defined in Sec.~\ref{sec:nhm}. The
   orthorhombic region is colored light and the tetragonal region is colored
   dark green.     
  \label{fig:gruppen}
}
 \end{figure}

In the present and the preceding paper~\cite{\lafeaso} we report evidence that
the correlated atomic-like motion of the electrons in magnetic or
superconducting bands produces the structural distortion, the
antiferromagnetic and the superconducting state in pure or doped LaFeAsO.  Our
findings shall be summarized in this section and are illustrated in the
schematic structural phase diagram shown in Fig.~\ref{fig:gruppen}.

In Sec.~\ref{sec:nonexistence} we have shown that undoped tetragonal LaFeAsO
does not possess a superconducting band. The symmetry of the space group
$P4/nmm$ produces band degeneracies which make it impossible to construct a
superconducting band in the band structure of undoped tetragonal
LaFeAsO. Thus, the electron system cannot gain the nonadiabatic condensation
energy $\Delta E$ by occupying an atomic-like state producing the spin-boson
interaction leading to superconductivity.

However, in Ref.~\cite{\lafeaso} we could show that undoped LaFeAsO possesses
a magnetic band related to the experimentally observed antiferromagnetic
structure. Thus, the system may lower its correlation energy by occupying an
atomic-like state producing the antiferromagnetic order as is experimentally
well-established ~\cite{clarina, nomura,kitao,nakai} below $137K$. This
magnetic structure is necessarily accompanied by a structural distortion of
the crystal with the space group $Pnn2$ going beyond the
magnetostriction. This distortion is realized by a displacement of the Fe and
O atoms which is also experimentally established~\cite{clarina}.

\begin{table}
\caption{Subgroup diagram of the space groups listed in Table~\ref{tab:gruppen}
\label{tab:diagramm}}
\begin{center}
\begin{tabular}[t]{ccc}
\hline
\multicolumn{3}{|c|}{$P4/nmm$}\\
\hline
$\downarrow$&&$\downarrow$\\
\begin{tabular}{|c|}\hline$Imma$\\\hline\end{tabular}&&
\begin{tabular}{|c|}\hline$P\overline{4}m2$\\\hline\end{tabular}
\\
$\downarrow$&&$\downarrow$\\
\begin{tabular}{|c|}\hline$Pnn2$\\\hline\end{tabular}&&
\begin{tabular}{|c|}\hline$Pmm2$\\\hline\end{tabular}
\\
\end{tabular}\hspace{1cm}
\end{center}
\ \\
\begin{flushleft}
Notes to Table~\ref{tab:diagramm}
\end{flushleft}
\begin{enumerate}
\item $P4/nmm$ is the space group of tetragonal paramagnetic LaFeAsO.
\item An arrow points to the subgroup of the group at the tail of the arrow. 
\item Neither $Imma$ nor $Pnn2$ is a subgroup of $P\overline{4}m2$. Consequently, 
  the doping destroys the antiferromagnetic order.
\item $Pmm2$ is a subgroup of $P\overline{4}m2$. Consequently, the
  non-magnetic orthorhombic distortion can survive in doped LaFeAsO.
\end{enumerate}
\end{table}


Nevertheless, the electron system in undoped LaFeAsO has another possibility
to gain $\Delta E$ in a {\em non-}magnetic state: the displacements of the Fe
and O atoms depicted in Fig.~\ref{fig:structures} (b) realize the orthorhombic
space group $Pmm2$ possessing a superconducting band and, consequently,
allowing an atomic-like state which produces spin-boson interaction. We
suppose that this spin-boson interaction is active between $155K$ and $137K$,
that is, in the non-magnetic region which is nonetheless orthorhombically
distorted. The spin-boson interaction would produce Cooper pairs below a
transition temperature $T_c$ which, however, does not happen. Before reaching
$T_c$ the system becomes antiferromagnetic at $137K$, see
Fig.~\ref{fig:gruppen}.

The displacements of the Fe and O atoms in the non-magnetic state and in the
antiferromagnetic state are depicted in Fig.~\ref{fig:structures} (b) of the
present paper and Fig.~3 (c) of Ref.~\cite{\lafeaso}, respectively. They are
almost identical, but there is a difference in $z$ direction: In the
non-magnetic system, the displacements realize the space group $Pmm2$ and are
periodic with $\bm T_3$ as given in Fig.~\ref{fig:structures} (b).  In the
antiferromagnetic system, on the other hand, they realize together with the
magnetic moments the space group $Pnn2$ and are no longer periodic with the
lattice vector $\bm T_3$ in the non-magnetic system, but with the lattice
vector $\bm T_3$ as given in Fig.~3 (c) of Ref.~\cite{\lafeaso} which is twice
as long as $\bm T_3$ in the non-magnetic system. Within the antiferromagnetic
unit cell, however, a translation by $\bm T_3/2$ effects an inversion of the
displacements.

The situation is changed in the doped system LaFeAsO$_{1-x}$F$_x$ because the
doping reduces the strong symmetry of the space group $P4/nmm$.  As
substantiated in Sec.~\ref{sec:spacegroup}, the tetragonal group
$P\overline{4}m2$ (115) is best adapted to the symmetry of the atomic-like
electrons in weakly doped LaFeAsO.  Indeed, the degeneracies within the
superconducting bands which suppress superconductivity in LaFeAsO are removed
in $P\overline{4}m2$ and the active band of LaFeAsO (as denoted in
Fig.~\ref{fig:bandstr} by the bold line) becomes a superconducting band. Thus,
the electron system now is able to lower its correlation energy by producing
the spin-boson interaction leading to superconductivity. This superconducting
state is in fact experimentally established~\cite{kamihara}.

The antiferromagnetic state cannot survive in the doped system because the
space group $Pnn2$ of the antiferromagnetic state is {\em not} a subgroup of
the space group $P\overline{4}m2$ of the atomic-like motion in the doped
material, see the subgroup diagram in Table~\ref{tab:diagramm}. As is
experimentally observed, the antiferromagnetism actually disappears at
relatively small doping levels between $x = 0.03$ and $0.05$~\cite{huang}. The
non-magnetic orthorhombically distorted state, on the other hand, may readily
survive in the doped system because the space group $Pmm2$ of the non-magnetic
distorted state is a subgroup of $P\overline{4}m2$. Also this phenomenon is
experimentally established: the orthorhombic structural distortion extends
beyond the antiferromagnetic phase and coexists with
superconductivity~\cite{huang}, see Fig.~\ref{fig:gruppen}.

Table~\ref{tab:gruppen} compiles all the space groups of LaFeAsO determined in
the present paper and in Ref.~\cite{\lafeaso} and Table~\ref{tab:diagramm}
illustrates the subgroup relations of these space groups.

\section{Conclusions}
\label{sec:conclusions}
In view of our present results on LaFeAsO and former results on the magnetic
materials Cr~\cite{ea}, Fe~\cite{ef}, La$_2$CuO$_4$~\cite{\joslacuo}, and
YBa$_2$Cu$_3$O$_6$~\cite{\ybacuo}, on the high-T$_c$ superconductors
La$_2$CuO$_4$~\cite{\joslacuo}, YBa$_2$Cu$_3$O$_7$, and
MgB$_2$~\cite{\josybacuobeide}, and on numerous elemental
superconductors~\cite{es2,\josm}, we propose that the nonadiabatic condensation
energy $\Delta E$ (see Sec.~\ref{sec:superconductivity}) generally
stabilizes magnetism and superconductivity. This proposition implies that the
mechanism of Cooper pair formation established within the NHM is generally
responsible for superconductivity.

A fundamental message of the NHM is that the constraining forces operating in
narrow, partly filled superconducting bands are {\em required} for the
Hamiltonian of the system to possess eigenstates in which the electrons form
Cooper pairs, see Sec.~\ref{sec:superconductivity}. This fundamental
statement is straightforwardly corroborated by the physical properties of
LaFeAsO which is non-superconducting in the space group $P4/nmm$, but becomes
superconducting once this symmetry is reduced by the doping, see
Sec.~\ref{sec:superconductingband}.

In this notation, the superconducting state in doped LaFeAsO is not caused by
the proximity of the system to the antiferromagnetic state in the undoped
material because both, the superconducting and the antiferromagnetic state,
are produced by different atomic-like states in the active band of LaFeAsO
which are mutually exclusive. The special properties of the superconducting
state are determined by the ``crystal-spin-1 bosons'' mediating Cooper pair
formation in a superconducting band \cite{\josybacuo,\josn}. These
crystal-spin-1 bosons are the energetically lowest boson excitations that
possess the crystal-spin angular momentum $1\cdot\hbar$ and are sufficiently
stable to transport it through the crystal, see
Sec.~\ref{sec:superconductivity}. We propose \cite{\josybacuo} that they are
coupled phonon-plasmon modes which have dominant phonon character in the
isotropic lattices of the transition elements and, hence, confirm the
electron-phonon mechanism that enters the BCS theory~\cite{bcs} in these
materials.  However, phonon-like excitations are not able to transport
crystal-spin angular-momenta within the two-dimensional layers of the
high-$T_c$ superconductors~\cite{ehtc}. Within the two-dimensional layers,
energetically higher lying excitations of dominant plasmon character,
i.e. crystal-spin-1 bosons cause Cooper pair formation instead.

\begin{acknowledgements}
We wish to thank Ernst Helmut Brandt for valuable discussion.
\end{acknowledgements}

%

\appendix

\renewcommand{\thetable}{\thesection.\arabic{table}}
\setcounter{table}{0}

\section{Definition of superconducting bands}
\label{sec:definition}
According to the definition in Ref.~\cite{\josybacuo}, an energy band of a
given material is called ``superconducting band'' if the Bloch functions of
this band 
\begin{itemize}
\item 
can be unitarily transformed into
\emph{spin-dependent} Wannier functions as defined by Eq.~(A22) of
Ref.~\cite{\enhm} which are
\begin{itemize}
\item centered on one sort of the atoms of the material;
\item symmetry-adapted to the (full) space group $G$ of this material; and
\item localized as well as possible~\cite{ew1};
\end{itemize}
\item and cannot be unitarily transformed into usual, i.e., spin-independent
  Wannier functions with these properties.
\end{itemize}

\FloatBarrier

\onecolumngrid

\section{Group-theoretical tables}

\label{sec:tables}

\begin{table}[b]
\caption{
Single- and double-valued representations of all the energy bands of
paramagnetic undoped LaFeAsO with symmetry-adapted and optimally  
localized (spin-dependent) Wannier functions centered at the Fe atoms. 
\label{tab:wf129}}
\begin{tabular}[t]{ccccccc}
\multicolumn{7}{c}{\em (a)\quad Single-valued representations}\\
& $\Gamma$ & $M$ & $Z$ & $A$ & $R$ & $X$\\
\hline
Band 1\ \ & $\Gamma^+_1$ + $\Gamma^-_4$ & $M_2$ & $Z^+_3$ + $Z^-_2$ & $A_4$ & $R_1$ & $X_1$\\
Band 2\ \ & $\Gamma^+_2$ + $\Gamma^-_3$ & $M_4$ & $Z^+_4$ + $Z^-_1$ & $A_2$ & $R_2$ & $X_2$\\
Band 3\ \ & $\Gamma^+_3$ + $\Gamma^-_2$ & $M_4$ & $Z^+_1$ + $Z^-_4$ & $A_2$ & $R_1$ & $X_1$\\
Band 4\ \ & $\Gamma^+_4$ + $\Gamma^-_1$ & $M_2$ & $Z^+_2$ + $Z^-_3$ & $A_4$ & $R_2$ & $X_2$\\
\hline\\
\end{tabular}\hspace{1cm}
\begin{tabular}[t]{ccccccc}
\multicolumn{7}{c}{\em (b)\quad Double-valued representations}\\
& $\Gamma$ & $M$ & $Z$ & $A$ & $R$ & $X$\\
\hline
Band 1\ \ & $\Gamma^+_6$ + $\Gamma^-_7$ & $M_5$ & $Z^+_7$ + $Z^-_6$ & $A_5$ & $R_3$ + $R_4$ & $X_3$ + $X_4$\\
Band 2\ \ & $\Gamma^+_7$ + $\Gamma^-_6$ & $M_5$ & $Z^+_6$ + $Z^-_7$ & $A_5$ & $R_3$ + $R_4$ & $X_3$ + $X_4$\\
\hline\\
\end{tabular}\hspace{1cm}
\ \\
\begin{flushleft}
Notes to Table~\ref{tab:wf129}
\end{flushleft}
\begin{enumerate}
\item Paramagnetic undoped LaFeAsO has the space group $P4/nmm =
  \Gamma_qD^{7}_{4h}$ (129).
\item The energy bands belonging to the (spin-dependent) Wannier functions centered
  at the La or As atoms are not listed. The two branches of all these
  bands are degenerate at points $M$ and $A$, too.
\item The notations of the single-valued representations (a) are given in
  Table A.1 of Ref.~\protect\cite{\lafeaso}.
\item The notations of the double-valued representations (b) may be identified
  from Table 6.13 in Ref.~\protect\cite{bc}.
\item Each row defines one band consisting of two branches, because there
  are two Fe atoms in the unit cell.
\item The bands are determined by Eq.~(23) of Ref.~\protect\cite{\joslacuo}. 
\item Assume a band of the symmetry in any row of this table to exist in the
  band structure of LaFeAsO.
  Then the Bloch functions of this band can be unitarily transformed into
  Wannier functions that are
\begin{itemize}
\item localized as well as possible; 
\item centered at the Fe atoms; and
\item symmetry-adapted to the space group $P4/nmm$.  
\end{itemize}
These Wannier function are usual (spin-independent) Wannier function if the
considered band is characterized by the single-valued representations (a). They
are spin-dependent if the band is characterized by the double-valued
representations (b).
\end{enumerate}
\end{table}


\begin{table}
  \caption{
    Character tables of the single-valued irreducible representations of the 
    tetragonal space group $P\overline{4}m2 = \Gamma_qD^5_{2d}$ (115).
  \label{tab:rep115}}
\begin{tabular}[t]{cccccc}
\\
$\Gamma (000)$, $M (\frac{1}{2}\frac{1}{2}0)$,  & $$ & $$ & $\{S^-_{4z}|\overline{\frac{1}{2}}00\}$ & $\{C_{2b}|000\}$ & $\{\sigma_y|0\frac{1}{2}0\}$\\
$Z (00\frac{1}{2})$, $A (\frac{1}{2}\frac{1}{2}\frac{1}{2})$ & $\{E|000\}$ & $\{C_{2z}|\overline{\frac{1}{2}}\frac{1}{2}0\}$ & $\{S^+_{4z}|0\frac{1}{2}0\}$ & $\{C_{2a}|\overline{\frac{1}{2}}\frac{1}{2}0\}$ & $\{\sigma_x|\overline{\frac{1}{2}}00\}$\\
\hline
$\Gamma_1$, $M_1$, $Z_1$, $A_1$ & 1 & 1 & 1 & 1 & 1\\
$\Gamma_2$, $M_2$, $Z_2$, $A_2$  & 1 & 1 & 1 & -1 & -1\\
$\Gamma_3$, $M_3$, $Z_3$, $A_3$  & 1 & 1 & -1 & 1 & -1\\
$\Gamma_4$, $M_4$, $Z_4$, $A_4$  & 1 & 1 & -1 & -1 & 1\\
$\Gamma_5$, $M_5$, $Z_5$, $A_5$  & 2 & -2 & 0 & 0 & 0\\
\hline\\
\end{tabular}\hspace{1cm}
\begin{tabular}[t]{ccccc}
$R (0\frac{1}{2}\frac{1}{2})$, $X (0\frac{1}{2}0)$ & $\{E|000\}$ & $\{C_{2z}|\overline{\frac{1}{2}}\frac{1}{2}0\}$ & $\{\sigma_y|0\frac{1}{2}0\}$ & $\{\sigma_x|\overline{\frac{1}{2}}00\}$\\
\hline
$R_1$, $X_1$ & 1 & 1 & 1 & 1\\
$R_2$, $X_2$ & 1 & -1 & 1 & -1\\
$R_3$, $X_3$ & 1 & 1 & -1 & -1\\
$R_4$, $X_4$ & 1 & -1 & -1 & 1\\
\hline\\
\end{tabular}\hspace{1cm}
\ \\
\begin{flushleft}
Notes to Table~\ref{tab:rep115}
\end{flushleft}
\begin{enumerate}
\item The group $P\overline{4}m2$ is optimally adapted to the symmetry of
    atomic-like electrons at the Fe sites in weakly doped LaFeAsO.
\item The character tables are determined from Table 5.7 in
  Ref.~\protect\cite{bc}, where the symmetry operations given in Table 5.7 of
  Ref.~\protect\cite{bc} are changed into the operations used in this paper as
  described in the notes to Table~\ref{tab:gruppen}.
\item The group $P\overline{4}m2$ is symmorphic. Hence, the point group
  operations are, on their own, symmetry operations of the crystal.
  Nevertheless, in this paper, the point group operations are associated with
  translations in consequence of the unusual position of our coordinate system
  given in Fig.~\ref{fig:structures} (a).
\end{enumerate}
\end{table}


\begin{table}[t]
\caption{Character tables of the double-valued irreducible representations of the
  space group $P\overline{4}m2$.   
    \label{tab:rep115Z}}
\begin{tabular}[t]{cccccccc}
\\
$\Gamma (000)$, & $$ & $$ & $$ & $$ & $$ & $\{\overline{C}_{2a}|\overline{\frac{1}{2}}\frac{1}{2}0\}$ & $\{\overline{\sigma}_y|0\frac{1}{2}0\}$\\
$M (\frac{1}{2}\frac{1}{2}0)$, & $$ & $$ & $$ & $$ & $$ & $\{\overline{C}_{2b}|000\}$ & $\{\sigma_x|\overline{\frac{1}{2}}00\}$\\
$Z (00\frac{1}{2})$, & $$ & $$ & $\{\overline{C}_{2z}|\overline{\frac{1}{2}}\frac{1}{2}0\}$ & $\{S^-_{4z}|\overline{\frac{1}{2}}00\}$ & $\{\overline{S}^+_{4z}|0\frac{1}{2}0\}$ & $\{C_{2b}|000\}$ & $\{\sigma_y|0\frac{1}{2}0\}$\\
$A (\frac{1}{2}\frac{1}{2}\frac{1}{2})$ & $\{E|000\}$ & $\{\overline{E}|000\}$ & $\{C_{2z}|\overline{\frac{1}{2}}\frac{1}{2}0\}$ & $\{S^+_{4z}|0\frac{1}{2}0\}$ & $\{\overline{S}^-_{4z}|\overline{\frac{1}{2}}00\}$ & $\{C_{2a}|\overline{\frac{1}{2}}\frac{1}{2}0\}$ & $\{\overline{\sigma}_x|\overline{\frac{1}{2}}00\}$\\
\hline
$\Gamma_6$, $M_6$, $Z_6$, $A_6$ & 2 & -2 & 0 & $\sqrt{2}$ & $-\sqrt{2}$ & 0 & 0\\
$\Gamma_7$, $M_7$, $Z_7$, $A_7$ & 2 & -2 & 0 & $-\sqrt{2}$ & $\sqrt{2}$ & 0 & 0\\
\hline\\
\end{tabular}\hspace{1cm}
\begin{tabular}[t]{cccccc}
\\
$R (0\frac{1}{2}\frac{1}{2})$, & $$ & $$ & $\{\overline{C}_{2z}|\overline{\frac{1}{2}}\frac{1}{2}0\}$ & $\{\overline{\sigma}_y|0\frac{1}{2}0\}$ & $\{\sigma_x|\overline{\frac{1}{2}}00\}$\\
$X (0\frac{1}{2}0)$ & $\{E|000\}$ & $\{\overline{E}|000\}$ & $\{C_{2z}|\overline{\frac{1}{2}}\frac{1}{2}0\}$ & $\{\sigma_y|0\frac{1}{2}0\}$ & $\{\overline{\sigma}_x|\overline{\frac{1}{2}}00\}$\\
\hline
$R_5$, $X_5$ & 2 & -2 & 0 & 0 & 0\\
\hline\\
\end{tabular}\hspace{1cm}
\ \\
\begin{flushleft}
Note to Table~\ref{tab:rep115Z}
\end{flushleft}
\begin{enumerate}
\item The character tables are determined from Table 6.13 in
  Ref.~\protect\cite{bc}, cf. the notes to Table~\ref{tab:rep115}. 
\end{enumerate}
\end{table}


\begin{table}[!]
\caption{
Compatibility relations between the space groups $P4/nmm$ and $P\overline{4}m2$.
\label{tab:faltena}
}
\begin{tabular}[t]{cccccccccc}
\hline\\
\end{tabular}\hspace{1cm}
\begin{tabular}[t]{cccccccccc}
\multicolumn{10}{c}{$\Gamma$, $Z$}\\
\hline
$R^+_1$ & $R^+_2$ & $R^+_3$ & $R^+_4$ & $R^+_5$ & $R^-_1$ & $R^-_2$ & $R^-_3$ & $R^-_4$ & $R^-_5$\\
$R_1$ & $R_2$ & $R_4$ & $R_3$ & $R_5$ & $R_3$ & $R_4$ & $R_2$ & $R_1$ & $R_5$\\
\hline\\
\end{tabular}\hspace{1cm}
\begin{tabular}[t]{cccc}
\multicolumn{4}{c}{$M$, $A$}\\
\hline
$R_1$ & $R_2$ & $R_3$ & $R_4$\\
$R_5$ & $R_1$ + $R_3$ & $R_5$ & $R_2$ + $R_4$\\
\hline\\
\end{tabular}\hspace{1cm}
\begin{tabular}[t]{cc}
\multicolumn{2}{c}{$R$, $X$}\\
\hline
$R_1$ & $R_2$\\
$R_1$ + $R_4$ & $R_3$ + $R_2$\\
\hline\\
\end{tabular}\hspace{1cm}
\ \\
\begin{flushleft}
Notes to Table~\ref{tab:faltena}
\end{flushleft}
\begin{enumerate}
\item The Brillouin zones for $P4/nmm$ and $P\overline{4}m2$ are identical. 
\item The letter $R$ stands for the letter denoting the relevant point of
  symmetry.  For example, at point M the representations $R_1, R_2, \ldots$
  stand for $M_1, M_2, \ldots$ .
\item The upper and lower rows list the representations of $P4/nmm$ and
  $P\overline{4}m2$, respectively.  
  The representations in the same column are compatible in the
  following sense: Bloch functions that are basis functions of a
  representation $D_i$ in the upper row can be unitarily transformed into
  the basis functions of the representation given below $D_i$.
\item The compatibility relations are determined in the way described in
  Ref.~\cite{eabf}.
\item The representations are labeled as given in Table A.1 of
  Ref.~\cite{\lafeaso} and Table~\ref{tab:rep115}, respectively.
\end{enumerate}
\end{table}


\begin{table}
\caption{
    Compatibility relations between the single-valued (upper row) and
    double-valued (lower row) representations of the space group
    $P\overline{4}m2$. 
\label{tab:faltenb}}
\begin{tabular}[t]{ccccc}
\multicolumn{5}{c}{$\Gamma$, $Z$, $M$, $A$}\\
\hline
$R_1$ & $R_2$ & $R_3$ & $R_4$ & $R_5$\\
$R_6$ & $R_6$ & $R_7$ & $R_7$ & $R_6$ + $R_7$\\
\hline\\
\end{tabular}\hspace{1cm}
\begin{tabular}[t]{cccc}
\multicolumn{4}{c}{$R$, $X$}\\
\hline
$R_1$ & $R_3$ & $R_2$ & $R_4$\\
$R_5$ & $R_5$ & $R_5$ & $R_5$\\
\hline\\
\end{tabular}\hspace{1cm}
\ \\
\begin{flushleft}
Notes to Table~\ref{tab:faltenb}
\end{flushleft}
\begin{enumerate}
\item The letter $R$ stands for the letter denoting the relevant point of
  symmetry.  For example, at point M the representations $R_1, R_2, \ldots$
  stand for $M_1, M_2, \ldots$ .
\item The single-valued and double-valued representations are listed in
  Tables~\ref{tab:rep115} and~\ref{tab:rep115Z}, respectively. 
\item Each column lists the double-valued representation $R_i\times D_{1/2}$
  below the single-valued representation $R_i$, where $D_{1/2}$ denotes the
  two-dimensional double-valued representation of the three-dimensional
  rotation group $O(3)$ given, e.g., in Table 6.1 of Ref.~\cite{bc}.
\end{enumerate}
\end{table}


\begin{table}[!]
\caption{
Single- and double-valued representations of all the energy bands in the tetragonal
space group $P\overline{4}m2 = \Gamma_qD^5_{2d}$ (115) with symmetry-adapted and
optimally localized (spin-dependent) Wannier functions centered at the Fe atoms. 
\label{tab:wf115}}
\begin{tabular}[t]{ccccccc}
\multicolumn{7}{c}{\em (a)\quad Single-valued representations}\\
 & $\Gamma$ & $M$ & $Z$ & $A$ & $R$ & $X$\\
\hline
Band 1\ \ & 2$\Gamma_1$ & $M_1$ + $M_3$ & 2$Z_4$ & $A_2$ + $A_4$ & $R_1$ + $R_4$ & $X_1$ + $X_4$\\
Band 2\ \ & 2$\Gamma_2$ & $M_2$ + $M_4$ & 2$Z_3$ & $A_1$ + $A_3$ & $R_3$ + $R_2$ & $X_3$ + $X_2$\\
Band 3\ \ & 2$\Gamma_3$ & $M_1$ + $M_3$ & 2$Z_2$ & $A_2$ + $A_4$ & $R_3$ + $R_2$ & $X_3$ + $X_2$\\
Band 4\ \ & 2$\Gamma_4$ & $M_2$ + $M_4$ & 2$Z_1$ & $A_1$ + $A_3$ & $R_1$ + $R_4$ & $X_1$ + $X_4$\\
\hline\\
\end{tabular}\hspace{1cm}
\begin{tabular}[t]{ccccccc}
\multicolumn{7}{c}{\em (b)\quad Double-valued representations}\\
 & $\Gamma$ & $M$ & $Z$ & $A$ & $R$ & $X$\\
\hline
Band 1\ \ & 2$\Gamma_6$ & $M_6$ + $M_7$ & 2$Z_7$ & $A_6$ + $A_7$ & 2$R_5$ & 2$X_5$\\
Band 2\ \ & 2$\Gamma_7$ & $M_6$ + $M_7$ & 2$Z_6$ & $A_6$ + $A_7$ & 2$R_5$ & 2$X_5$\\
\hline\\
\end{tabular}\hspace{1cm}
\ \\
\begin{flushleft}
Notes to Table~\ref{tab:wf115}
\end{flushleft}
\begin{enumerate}
\item The notations of the single-valued and double-valued representations are
  given in Tables~\ref{tab:rep115} and~\ref{tab:rep115Z}, respectively.
\item Each row defines one band consisting of two branches, because there
  are two Fe atoms in the unit cell.
\item The bands are determined by Eq.~(23) of Ref.~\protect\cite{\joslacuo}. 
\item Assume a band of the symmetry in any row of this table to exist in the
  band structure of LaFeAsO.
  Then the Bloch functions of this band can be unitarily transformed into
  Wannier functions that are
\begin{itemize}
\item localized as well as possible; 
\item centered at the Fe atoms; and
\item symmetry-adapted to the space group $P\overline{4}m2$.
\end{itemize}
These Wannier function are usual (spin-independent) Wannier function if the
considered band is characterized by the single-valued representations
(a). They are spin-dependent if the band is characterized by the double-valued
representations (b).
\end{enumerate}
\end{table}


\end{document}